\UseRawInputEncoding
\documentclass[prd,aps,preprint,amsmath,nofootinbib,amssymb,eqsecnum,showkeys,tightenlines]{revtex4-1}

\usepackage{amsmath, latexsym, amssymb, hyperref, graphicx, color, multirow, makecell, diagbox}
\usepackage[table]{xcolor}
\usepackage{tabularx}
\usepackage[T1]{fontenc}
\usepackage[capitalize]{cleveref}
\usepackage{multirow}
\usepackage{slashed}
\numberwithin{equation}{section}
\usepackage{verbatim}
\allowdisplaybreaks

\def\nnb{\nonumber}

\newcommand{\be}{\begin{equation}}
\newcommand{\ee}{\end{equation}}
\newcommand{\bea}{\begin{eqnarray}}
\newcommand{\eea}{\end{eqnarray}}
\newcommand{\ba}{\begin{array}}
\newcommand{\ea}{\end{array}}

\def\be{\begin{equation}}
\def\ee{\end{equation}}
\def\bea{\begin{eqnarray}}
\def\eea{\end{eqnarray}}
\def\nnb{\nonumber}
\def\xslash\sharp1{{\rlap{$\sharp1$}/}}

\begin{document}
\title{Majorana CP violating phases }
\author{Chao-Shang Huang}
\email{csh@itp.ac.cn}
\affiliation{
\normalsize CAS Key Laboratory of Theoretical Physics, Institute of Theoretical Physics, \\
Chinese Academy of Sciences, Beijing 100190, China\\
}

\begin{center}
\begin{abstract}
The two Majorana CP-violating phases can not be determined by experiments on neutrino oscillations. It is difficult even almost impossible to measure two Majorana CP-violating phases since they are only sensitive to lepton-number-violating processes. One must take some assumption on the structure of neutrino mass matrix and their flavor mixing mechanism hidden behind in phenomenological models in order to determine Majorana CP-violating phases. Two models on the symmetry of the neutrino mass matrix are proposed in this paper. The Majorana CP-violating phases and the effective Majorana neutrino mass $ m_{ee}$ are computed in the two models. Using data in experiments on neutrino oscillations and using the limit of the absolute neutrino mass scale which is from cosmological observations, the numerical values of the Majorana CP-violating phases and the effective Majorana neutrino mass $ m_{ee}$ are obtained.
\end{abstract}
\end{center}

\maketitle

\section{introduction}\label{sec1}

In recent years much progress has been made in experiments on neutrino oscillations. However, the
absolute neutrino mass scale and two Majorana cp-violating phases can not be determined by experiments on neutrino oscillations. The absolute neutrino mass scale can be determined or constrained by cosmological observations or experiments on double-beta decays without neutrinos ($0 \nu 2 \beta$). At the present stage it is difficult even almost impossible to measure two Majorana CP-violating phases since they are only sensitive to lepton-number-violating processes\footnote{Although in resent years some lepton-flavor-violating processes such as $l_i \rightarrow l_j \gamma\gamma$ and $l_i \rightarrow l_j \gamma$ have been measured by experiments and the upper limit on BR $(90\% ~\text{c.l.})$ is $10^{-8}-10^{-13}$ for $l_i \rightarrow l_j \gamma$ and $10^{-4}-10^{-11}$ for $ l_i \rightarrow l_j \gamma\gamma$ \cite{lfv1,lfv} only, the lepton-number-violating processes have not been seen till now.}. One must take some assumption on the structure of neutrino mass matrix and their flavor mixing mechanism hidden behind in phenomenological models in order to determine two Majorana CP-violating phases \cite{phemo}.
Some new techniques for producing and measuring neutrinos and antineutrinos will probably be developed in the future \cite{df}, and then experiments on neutrino-antineutrino oscillations will make remarkable progress and the Majorana CP-violating phases will be measured. Then, verification of models should be feasible by using the experimentally measured values of two Majorana CP-violating phases.

As early as the beginning of the this century, according to results from the SNO ( Sudbury solar neutrino experiment) and the K2K (KEK to Kamioka) long-baseline neutrino experiment, the combined existing data on neutrino oscillations, P. F. Harrison et al. present the tri-bimaximal mixing \cite{hps} of the PMNS (Pontecorvo B, Maki Z, Nakagawa M and Sakata S) mixing matrix as
\bea
U_\text{TBM} =\left(\begin{array}{ccc}
  2/ \sqrt{6}   & 1/ \sqrt{3}  &   0 \\
 -1/ \sqrt{6}   & 1/ \sqrt{3}  & 1/ \sqrt{2}  \\
 -1/\sqrt{6}  & 1/ \sqrt{3} & -1/ \sqrt{2}
\end{array}\right).\label{utbm}
\eea
As of August 2005, the three neutrino mixing parameters have been established, all at
95\% c.l., based on a global fit on neutrino oscillation experiments as reviewed in \cite{flmp}. They are consistent with the tri-bimaximal mixing. Experimental data show that the atmospheric neutrino mixing angle $\theta_{23}$ is close to the maximum. There are many works which attempt to explain largeness of $\theta_{23}$ \cite{gl}. Babu, Ma, and Valle (BMV) \cite{bmv,ma} suggest a Majorana mass matrix as
\bea
M_{\nu}=\left(\begin{array}{ccc}
a & r & r^* \\
r & s & b \\
r^* & b & s^*
\end{array}\right),
\eea
where $r$ and $s$ are in general complex while $a$ and $b$ remain real and it is found that this $M_{\nu}$ yields maximal atmospheric mixing and a non-zero $U_{e3}$ as soon as $r$ and $s$ are complex. Eq.~(\ref{utbm}) shows that $|U_{\mu i}|=|U_{\tau i}|$ (for $i=1,2,3$) known as the $\mu-\tau$ reflection (or interchange ) symmetry\,\cite{hs,zzx1,rt1}. It is proved that the mass matrix of BMV always yields the $\mu-\tau$ reflection symmetry  and, as a consequence, exact maximal atmospheric neutrino mixing $(s_{23}=\pm c_{23})$ and maximal CP violation \cite{gl}.

It is noted \cite{csl} that the neutrino mass matrix
\bea
M_\text{TBM}=U^T_\text{TBM} D U_\text{TBM}~,
\eea
where $D= \text{Diag}(m_1, m_2, m_3)$ with $m_i~(i=1,2,3)$ being the neutrino masses, $M_\text{TBM}$ is 2-3 symmetric and magic. The magic neutrino mass matrix has advantage to examine symmetries of the neutrino mixing matrix and its phenomenological implications have been studied in the literature \cite{csl,lam,mm}. Within the framework of seesaw  mechanism and modula (for example, A4) flavor symmetry the magic neutrino mass matrix can be obtained \cite{mm}. Although the experimentally measured neutrino mixing matrix does not lead to a magical neutrino mass matrix, in real nature world it would be very probable that the neutrino mass matrix is magical and the mixing matrix is experimentally measured. This inspires us to present the first phenomenological model.

There are no the Majorana phases in $U_\text{TBM}$. In order to examine the Majorana phases a generalized tri-bimaximal mixing, $U_m= U_\text{TBM} P$ and $P=\text{Diag}(1, e^{i\xi_1}, e^{i\xi_2})$ with $\xi_1$ and $\xi_2$ being the Majorana phases, has been introduced \cite{ar} such that the
generalized tri-bimaximal mixing leads to the neutrino mass matrix
\bea
M_\text{TBM}=U_m D U^T_m~.
\eea
The generalized tri-bimaximal mixing enlightens us to propose the second phenomenological model.

In this paper we determine two Majorana CP-violating phases, by means of some assumption on the structure of neutrino mass matrix and their flavor mixing mechanism hidden behind in phenomenological models, and obtain their numerical results using data in experiments on neutrino oscillations and the limit of the absolute neutrino mass scale determined due to cosmological observations.

In Section \ref{sec2}, we calculate the Majorana CP-violating phases as well as the effective Majorana neutrino masses $ m_{ee}$ by means of the assumption in the first model that the Majorana neutrino mass matrix is magic and the mixing matrix is experimentally measured in real nature world. We review the $\mu-\tau$ reflection symmetry and its limitations simply in Section \ref{sec3}. In Section \ref{sec4}, we propose the second model based on the assumption about the symmetry of the neutrino mass matrix and compute the Majorana CP-violating phases as well as the effective Majorana neutrino masses $m_{ee}$ in the model.  Finally we present summary and outlook in Section \ref{sec5}.

\section{The Magic Symmetry}\label{sec2}

A $n \times n$ matrix will be called magic if the row sums and the column sums are all
equal to a same common number \cite{csl}. In the flavoured basis, where the charged lepton mass matrix is diagonal, the magic Majorana neutrino mass matrix can be expressed as
\bea
m_\nu=\left(\begin{array}{ccc}
a & b & c \\
b & d & a+c-d \\
c & a+c-d & b-c+d
\end{array}\right)~.\label{mnu}
\eea
Since
\bea
m_{\alpha \beta} \equiv \sum_{i=1}^3 m_i U_{\alpha i} U_{\beta i}~,~~~~~\alpha,\beta=e, \mu, \tau,\label{mab}
\eea  with the neutrino mixing matrix $U=U_\text{PMNS}$\,\cite{pmns}, according to the assumption in our first model, one has
 \bea
 a &=& m_{ee}=m_1 c_{12}^2 c_{13}^2  e^{i\xi_1} + m_2 s_{12}^2 c_{13}^2  e^{i\xi_2} + m_3 s_{13}^2 e^{-2 i\delta}~,\nnb\\
 b &=& m_{e\mu}= -m_1 c_{12} c_{13} (s_{12} c_{23} + c_{12} s_{23} s_{13} e^{i\delta}) e^{i\xi_1} +m_2 s_{12} c_{13} (c_{12} c_{23} - s_{12} s_{23} s_{13} e^{i\delta}) e^{i\xi_2} +m_3  c_{13}  s_{23}  s_{13}  e^{-i\delta}~,\nnb\\
 c &=& m_{e\tau}= m_1  c_{12}  c_{13} (s_{12}  s_{23} - c_{12}  c_{23}  s_{13}  e^{i\delta}) e^{i\xi_1} - m_2  s_{12}  c_{13} (c_{12}  s_{23} + s_{12}  c_{23}  s_{13}   e^{i\delta})  e^{i\xi_2} +m_3  c_{13}  c_{23}  s_{13}  e^{-i\delta}~,\nnb\\
 d &=& m_{\mu\mu}= m_1( s_{12} c_{23} + c_{12} s_{23} s_{13} e^{i\delta})^2 e^{i\xi_1} + m_2 (c_{12} c_{23}- s_{12} s_{23} s_{13} e^{i\delta})^2 e^{i\xi_2} + m_3 c_{13}^2 s_{23}^2~,
 \eea
 where $\xi_1$, $\xi_2 $ and $\delta$ are two Majorana CP-violating phases and Dirac  CP-violating phase, respectively. It is straightforward from Eqs.~(\ref{mnu}, \ref{mab}) to obtain  $m_{\mu \tau}=a+c-d$ and $m_{\tau \tau}=b-c+d$. Solving them, we obtain
 \bea \label{e_11}
 e^{i\xi_1} & =& (m_3 (-s_{12} (c_{23} - s_{23}) (c_{13}^2 e^{- i\delta} s_{13} + e^{-2 i\delta}  e^{i\delta} s_{13}^3 +
         c_{13}^3 (c_{23} + s_{23}) + c_{13} s_{13}^2 (c_{23} + s_{23})) \nnb\\
 & & + c_{12} (-2 c_{13} c_{23} e^{- i\delta} s_{13} s_{23} - ed s_{13}^2 (c_{23} + s_{23}) \nnb\\
 & & + c_{13}^2 (c_{23}^3 + c_{23}^2 s_{23} + c_{23} s_{23}^2 + s_{23}^3))))/(m_1 (c_{13} (c_{23} + s_{23})\nnb\\
 & & + e^{i\delta} s_{13}) (s_{12} (-c_{23} + s_{23}) + c_{12} (c_{13} - e^{i\delta} s_{13} (c_{23} + s_{23}))))~,
\eea
\bea \label{e_21}
e^{i\xi_2} & =& (m_3 (-c_{13}^3 (c_{23} (-1 + s_{12}^2 - c_{12} e^{i\delta} s_{12} s_{13}) + (-1 + s_{12}^2 +c_{12} e^{i\delta} s_{12} s_{13}) s_{23}) (c_{23}^2 - s_{23}^2) \nnb\\
& & + c_{13}^2 ((e^{i\delta} + e^{- i\delta} - e^{- i\delta} s_{12}^2) s_{13} (c_{23}^2 - s_{23}^2)\nnb\\
& & + c_{12}^2 e^{i\delta} s_{13} (-c_{23}^4 + s_{23}^4) + c_{12} s_{12} (c_{23}^4 + c_{23}^2  s_{13}^2 \nnb\\
& & + 2 c_{23}^3 s_{23} - 2 c_{23}  s_{13}^2 s_{23} + 2 c_{23}^2 s_{23}^2 +  s_{13}^2 s_{23}^2 \nnb\\
& & + 2 c_{23} s_{23}^3 + s_{23}^4)) + ed s_{13}^2 (c_{12}^2 e^{i\delta} s_{13} (c_{23}^2 - s_{23}^2) \nnb\\
& & + e^{i\delta} s_{12}^2 s_{13} (-c_{23}^2 + s_{23}^2) + c_{12} s_{12} (c_{23}^2 (-1 + e^{2 i\delta} s_{13}^2) \nnb\\
& & + (-1 + e^{2 i\delta} s_{13}^2) s_{23}^2 - 2 c_{23} (s_{23} + e^{2 i\delta} s_{13}^2 s_{23})))  \nnb\\
& & + c_{13} e^{- i\delta} s_{13} (c_{23}^3 e^{i\delta} s_{12} s_{13} (-s_{12} + c_{12} e^{i\delta} s_{13}) + c_{23}^2 (2 c_{12}^2 e^{i\delta} s_{13} \nnb\\
& & - e^{i\delta} s_{12}^2 s_{13} - c_{12} s_{12} (2 + e^{2 i\delta} s_{13}^2)) s_{23} + e^{i\delta} s_{13} s_{23} (-1 + s_{12}^2 s_{23}^2 \nnb\\
& & + c_{12} e^{i\delta} s_{12} s_{13} s_{23}^2) + c_{23} (-2 c_{12} s_{12} s_{23}^2 - c_{12} e^{2 i\delta} s_{12} s_{13}^2 s_{23}^2 \nnb\\
& & + e^{i\delta} s_{13} (1 - 2 c_{12}^2 s_{23}^2 + s_{12}^2 s_{23}^2)))))/(m_2 (e^{i\delta} s_{12} s_{13} (c_{23} - s_{23})\nnb\\
& &  + c_{12} (c_{23} + s_{23})) (e^{i\delta} s_{13} + c_{13} (c_{23} + s_{23})) (c_{12} (c_{23} - s_{23})  \nnb\\
& & + s_{12} (c_{13} - e^{i\delta} s_{13} (c_{23} + s_{23}))))~.
\eea

The newest data in experiments on neutrino oscillations are given in TABLE II in Ref.~\cite{data}. We divide them as four sets as follows:
\begin{itemize}
    \item Without the addition of tabulated sk-atm $(\delta \chi)^2$ data:
    \\1) No (normal order of neutrino masses $m_1, m_2, m_3$)
    \\2) Io (inverse order of neutrino masses $m_1,m_2,m_3)$
    \item  With the addition of tabulated sk-atm $(\delta \chi)^2$ data:
    \\3) No
    \\4) Io
\end{itemize}
For simplicity, let us define op=$(s_{12}, s_{23}, s_{13}, c_{12}, c_{23}, c_{13}$, $\delta$, $\delta m_{21}^2$, $\delta m_{31}^2$). Then, the above four sets can be simply presented as   op1, op2, op3 and op4, respectively. We use the best fit values in numerical computations and the unit of mass is eV in the paper. To input the newest data in experiments on neutrino oscillations, Eqs.~(\ref{e_11}, \ref{e_21}) reduce to
\bea   \label{ope_1}
e^{i\xi_1}&=& (1.000-0.023 i) m_3)/m_1~,\nnb\\
e^{i\xi_2}&=& ((0.999 - 0.040 i) m_3)/m_2 \eea
for op1,
\bea  \label{ope_2}
e^{i\xi_1}&=& ((0.982 - 0.176 i) m_3)/m_1~,\nnb\\
e^{i\xi_2}&=& ((0.954 - 0.287 i) m_3)/m_2 \eea
for op2,
\bea  \label{ope3}
e^{i\xi_1}&=& ((0.990 - 0.157 i) m_3)/m_1~,\nnb\\
e^{i\xi_2}&=& ((0.995 - 0.116 i) m_3)/m_2 \eea
for op3, and
\bea  \label{ope4}
e^{i\xi_1}&=& ((0.984 - 0.166 i) m_3)/m_1~,\nnb\\
e^{i\xi_2}&=& ((0.962 - 0.259 i) m_3)/m_2 \eea
for op4.

From Eqs.~(\ref{ope_1}-\ref{ope4}), it is straightforward to obtain two Majorana CP-violating phases $\xi_1$ and $\xi_2$ and results are listed in TABLE \ref{xi12}~. It is deserved to emphasize that in this model the Majorana CP-violating phases are independent of neutrino masses $m_1$, $m_2$, and  $m_3$ indeed considering the modulus of $e^{i\xi_i}~(i=1,2)$ is 1.
\begin{table}[htb]
\begin{center}
\begin{tabular}{ccccc}
\hline
op &  op1  & op2 & op3 & op4 \\
\hline
\hline
$\xi_1/^\circ$ &~ 0.657   &~5.714  &~ 3.512   &~ 5.560 \\
\hline
\hline
$\xi_2/^\circ $ &~ 2.336   &~ 37.64 &~4.880   &~7.272 \\
\hline
\end{tabular}
\end{center}
\caption{The Majorana CP-violating phases}\label{xi12}
\end{table}

We now compute the absolute neutrino mass scale constrained by cosmological observations. We define $sm=m_1+m_2+m_3$. For No, $(m_1, m_2, m_3) = (m_1, (\delta m^2_{21} + m_1^2)^{1/2}, (\delta m^2_{31} + m_1^2)^{1/2}) $.  The upper limit of $sm$, 0.09 eV\,\cite{valpal}, leads to ($m_1,m_2,m_3$)=(0.01745,~0.01946,~0.05310) eV for op1 and ($m_1,m_2,m_3$)=(0.01747, 0.01948, 0.05306) eV for op3. For Io, $(m_1, m_2, m_3) = (m_1, (\delta m^2_{21} + m_1^2)^{1/2}, (\delta m^2_{32} + m_2^2)^{1/2})$ due to negative $\delta m^2_{32}$, there are no solutions for $m_i~(i=1,2,3)$ and there are still no solutions of $m_1, m_2$ even for $m_3=0$.  Therefore, we take $sm=0.12$ eV \cite{agh} and obtain ($m_3,m_2,m_1$)=(0.01585, 0.05243, 0.05172) eV for op2 and ($m_3,m_2,m_1$)=(0.01594, 0.05238, 0.05167) for op4.

Have known the Majorana CP-violating phases and neutrino masses ($m_1, m_2, m_3$), it is straightforward to compute the effective Majorana neutrino masses $ m_{ee}$ of which the absolute value can be measured in the neutrinoless double beta decay ($0\nu 2\beta$) experimentals
\be
m_{ee}= m_1 c^2_{12} c^2_{13} e^{i \xi_1} + m_2 s^2_{12} c^2_{13} e^{i \xi_2} + m_3 s^2_{13} e^{-2 i\delta}.
\ee
Results are listed in TABLE \ref{so10entriessu5u1x}. From TABLE \ref{so10entriessu5u1x}, predicted $|m_{ee}|/10^{-2}$  is smaller than or equal to 5.047 in all cases and is in agreement with the upper limit, 0.061-0.165, obtained  in the experiments \cite{agando}.
\begin{table}[htb]
\begin{center}
{\footnotesize
\begin{tabular}{ccc|ccccc}
  \hline
 $sm/$eV &  \multicolumn{2}{c|}{$0.09$} &  \multicolumn{4}{c}{0.12} \\
 \hline
 \hline
op & op1  & op3 & op1  & op2 & op3 & op4 \\
\hline
\hline
$m_{ee}/(0.01~\text{eV})$ &~1.869 - 0.0044 i &~1.742 - 0.0075 i &~3.097 - 0.0395 i &~5.047 + 0.05932 i &~2.954 -0.0500 i &~5.009 - 0.0092 i \\
\hline
\hline
$|m_{ee}|/(0.01~\text{eV})$  &~1.870   &~1.744    &~3.098   &~5.047 &~2.954   &~5.009 \\
\hline
    \end{tabular}}
    \end{center}
    \caption{The effective Majorana neutrino masses}\label{so10entriessu5u1x}
\end{table}

\section{The $\mu-\tau$ reflect symmetry}\label{sec3}
As pointed in Section \ref{sec1}, the tri-bimaximal mixing shows that $|U_{\mu i}|=|U_{\tau i}|$
(for i=1,2,3) known as the $\mu-\tau$ reflection symmetry. The $\mu-\tau$ reflection symmetry not only predicts $\theta_{23} = \pi/4$ and $\delta = \pm \pi/2$ but also constrains the corresponding Majorana phases to be 0 or $\pi/2$ \cite{zzx1}. The breaking of the $\mu-\tau$ reflection symmetry is also investigated in Ref.~\cite{rt1,zzx1}. In 2016, $s_{13}$ has been measured by the Daya Bay Reactor Neutrino Experiment \cite{dbr} with the result:
\begin{align*}
    \sin^2(2 \theta_{13})=0.0841\pm 0.0027(stat)\pm 0.0019(syst)~.
\end{align*}
In 2018, the RENO experiment reported \cite{RENO:2018dro}
\begin{align*}
    \sin^2(2 \theta_{13})=0.0896\pm 0.0048(stat)\pm 0.0047(syst)~.
\end{align*}
Therefore, the tri-bimaximal mixing should be revised to agree with the data. An interesting revision is the Co-Bimaximal Mixing (CBM) symmetry \cite{cbm,fmty,mty,hs1,ma1} of the PMNS mixing matrix which is obtained by taking $\theta_{23} = \pi/4$ and $ \delta = \pi/2 $ in $U_\text{PMNS}$, in addition to the Majorana phases $\xi_{1,2} = 0$. Therefore, $ \theta_{13}$ and $\theta_{12}$ in $U_\text{CBM}$ are free and can take the experimental values. That is,
\bea
U_\text{CBM} =\left(\begin{array}{ccc}
  c_{12} c_{13}  & s_{12} c_{13} &   -i s_{13} \\
(-1/\sqrt{2} ) (s_{12} - i  c_{12} s_{13})  & (1/\sqrt{2}) (c_{12} + i  s_{12} s_{13}) & -c_{13}/\sqrt{2}  \\
 (-1/\sqrt{2} ) (s_{12} + i  c_{12} s_{13})  & (1/\sqrt{2}) (c_{12} - i  s_{12} s_{13}) & c_{13}/\sqrt{2}
\end{array}\right)~.\label{ucbm}
\eea
 It is easy to know that the CBM mixing matrix is $\mu-\tau$ reflect symmetric. So, the neutrino mass matrix
\bea
M_\text{CBM}=U^T_\text{CBM} D U_\text{CBM} \label{mcbm}
\eea
is the same as the mass matrix of BMV, Eq.~(\ref{mnu}). The effects of the Majorana phases on $M_\text{CBM}$ are examed, and the different combinations of the Majorana phases consistent with small deviations from the $M_\text{CBM}$,  preserving the CBM symmetry in the PMNS matrix, have been found in Ref.~\cite{cbm}.

The newest data in experiments on neutrino oscillations\,\cite{data} show that $|U_{\mu i}|\approx |U_{\tau i}|$, i.e., the $\mu-\tau$ reflection symmetry is valid approximately so that the CBM mixing matrix, Eq.~(\ref{ucbm}), and the neutrino mass matrix $M_\text{CBM}$, Eq.~(\ref{mcbm}), are also valid approximately.

\section{The second model}\label{sec4}

\subsection{The assumption on the symmetry of the neutrino mass matrix}

There are no the Majorana phases in $U_\text{TBM}$ given by Eq.~(\ref{utbm}). In order to examine the Majorana phases a generalized tri-bimaximal mixing, $U_m = U_\text{TBM} P$ with $P = \text{Diag}(1, e^{i\xi_1}, e^{i\xi_2})$ and $\xi_1$ and $\xi_2$ being the Majorana phases, has been introduced \cite{ar} such that the generalized tri-bimaximal mixing leads to the neutrino mass matrix
\bea
M_\text{TBM}=U_m D U^T_m~,\label{mtbm}
\eea
which is magic. We know from the previous section that the CBM mixing matrix is $\mu-\tau$ reflect symmetric and leads to
\bea
M_\text{CBM}=U^T_\text{CBM} D U_\text{CBM}~.\label{mcbm}
\eea
The $\mu-\tau$ reflect symmetric neutrino mass matrix $M_\text{CBM}$ has been given by Eq.~(\ref{mnu}). We assume that the neutrino mass matrix is both magic and $\mu-\tau$ reflect symmetric. Then we get from Eqs.~(\ref{mtbm},~\ref{mcbm}) that
\bea
M_{\nu} U^{-1}=U^T M_{\nu}~,
\eea
where \bea
U= U_m U_c^{-1}~,~~~~U_c=U_ {CBM}~.
\eea
We compute the matrix $U$ first and it is straightforward, by means of Eqs.~(\ref{utbm}) and (\ref{ucbm}), to get
\bea
U =\left(\begin{array}{ccc}
U_{e1}& U_{e2}& U_{e3}\\ U_{\mu 1}& U_{\mu 2}& U_{\mu 3}\\ U_{\tau 1}& U_{\tau 2}& U_{\tau 3}
\end{array}\right)~,\label{uu}
\eea
where
\begin{align}
   U_{e1}&=(c_{13} (\sqrt{2} c_{12} + s_{12} e^{i\xi_1}))/\sqrt{3}~,\nonumber\\
   U_{e2}&=3/\sqrt{6}  + s_{12} c_{13} e^{i\xi_1}/\sqrt{3} +i s_{13} e^{i\xi_2}/\sqrt{2}~,\nonumber\\
   U_{e3}&=-c_{12} c_{13}/\sqrt{6} + s_{12} c_{13} e^{i\xi_1}/\sqrt{3} -i s_{13} e^{i\xi_2}/\sqrt{2}~,\nonumber\\
   U_{\mu 1}&=(-1/\sqrt{3} (s_{12} - i  c_{12} s_{13})+(((c_{12} + i s_{12} s_{13}) e^{i\xi_1})/\sqrt{6}~,\nonumber\\
   U_{\mu 2}&=1/6 (\sqrt{3} (s_{12} - i c_{12} s_{13}) + \sqrt{6} (c_{12} + i s_{12} s_{13}) e^{i\xi_1} + 3 c_{13} e^{i\xi_2})~,\nonumber\\
   U_{\mu 3}&=1/6 (\sqrt{3} (s_{12} - i c_{12} s_{13}) + \sqrt{6} (c_{12} + i s_{12} s_{13}) e^{i\xi_1} - 3 c_{13} e^{i\xi_2})~,\nonumber\\
   U_{\tau 1}&=(-1/\sqrt{3} (s_{12} - i  c_{12} s_{13})+(((c_{12} + i s_{12} s_{13}) e^{i\xi_1})/\sqrt{6}~,\nonumber\\
   U_{\tau 2}&=1/6 (\sqrt{3} (s_{12} - i c_{12} s_{13}) + \sqrt{6} (c_{12} + i s_{12} s_{13}) e^{i\xi_1} + 3 c_{13} e^{i\xi_2})~,\nonumber\\
   U_{\tau 3}&= 1/6 (\sqrt{3}(s_{12} + i  c_{12} s_{13})+\sqrt{6}(c_{12} - i  s_{12} s_{13}) e^{i\xi_1}+3 c_{13} e^{i\xi_2})~.
\end{align}
All of them are the function of $\xi_1, \xi_2$, as they should be.

The neutrino mass matrix is defined as Eq.~(\ref{mab}) which have been given in Section \ref{sec2}. Substituting the elements of the matrix $U$, i.e., Eq.~(\ref{uu}), into Eq.~(\ref{mab}), we have
\bea
    a &=& m_{ee}=m_1 U_{e1}^2+m_2 U_{e2}^2+m_3 U_{e3}^2~,\nonumber\\
    r &=& m_{e\mu}=m_1 U_{e1} U_{\mu 1}+m_2 U_{e2} U_{\mu 2}+m_3 U_{e3} U_{\mu 3}~,\nonumber\\ r^* &=& m_{e\tau}=m_1 U_{e1} U_{\tau 1} + m_2 U_{e2} U_{\tau 2}+m_3
U_{e3} U_{\tau 3}~,\nonumber\\
s &=& m_{\mu\mu}=m_1 U_{\mu 1}^2+m_2 U_{\mu 2}^2+m_3 U_{\mu 3}^2~,\nonumber\\
b &=&  m_{\mu \tau}=m_1 U_{\mu 1} U_{\tau 1}+m_2 U_{\mu 2} U_{\tau 2}+m_3 U_{\mu 3} U_{\tau 3}~,\nonumber\\
s^* &=& m_{\tau\tau}=m_1 U_{\tau 1}^2+m_2 U_{\tau 2}^2+m_3 U_{\tau 3}^2~, \label{many}
\eea
where we use the notation, $a,~b,~r,~s$ etc., same as those in Eq.~(\ref{mnu}). The assumption that neutrino mass matrix $M_{\nu}$ is $\mu-\tau$ reflect symmetric leads to $a$ and $b$ are real. And the assumption that $M_{\nu}$ is magic leads to
\be
s=a+r^*-b ~,\label{sa}
\ee
so that $M_{\nu}$ has four real parameters only.

\subsection{Majorana CP vialating phases}
From Eq.~(\ref{many}), it follows that
\bea
m_{\mu\mu}=m_{\tau\tau}^*~.
\eea
$a$ and $b$ are real means the constraints
\bea
Im[a]=Im[b]=0~.\label{im0}
\eea
For convenience let $e^{i\xi_1}$= $w_1$ = rw1 + i iw1, $e^{i\xi_2}$=$w_2$ = rw2 + i iw2 with rwj and iwj (j=1,2) being real. The constraints Eq.~(\ref{im0}) leads to  that  $w_1$ and $w_2$ are not completely independent of each other.
One obtains from the Eqs.~(\ref{sa}-\ref{im0}) that 
rw2 satisfies the following equation:
\bea
 k0 + k1\, rw2 + k2 \,rw2^2 + k3\, rw2^3 + k4\, rw2^4 + k5\, rw2^5 + k6\, rw2^6=0~, \label{cosxi2}
 \eea
 where
 \bea
 k_0 &=& -B_0 D_0 + A_0 C_0~, \nnb\\
 k_1 &=& (-B_1 D_0 - B_0 D_1 + A_0 C_1)~,\nnb\\
 k_2 &=& (-B_2 D_0 - B_1 D_1 -B_0 D_2 + A_1 C_1 +A_0 C_2)~,\nnb\\
 k_3 &=& (-B_2 D_1 - B_1 D_2 -B_0 D_3 + A_2 C_1 + A_1 C_2 +A_0 C_3)~,\nnb\\
 k_4 &=& (-B_2 D_2 - B_1 D_3 -B_0 D_4 + A_2 C_2 + A_1 C_3 +A_0 C_4)~,\nnb\\
 k_5 &=& (-B_2 D_3 - B_1 D_4 +A_2 C_3 + A_1 C_4)~,\nnb\\
 k_6 &=& (-B_2 D_4 + A_2 C_4)~.
 \eea
The $A_i$, $B_i~(i=0,1,2)$ and $C_j$, $D_j~(j=0,1,...,4)$ are listed in Appendix \ref{appa}.

\begin{table}[htb]
\begin{center}
\begin{tabular}{ccc|ccccc}
  \hline
 $sm/$eV &  \multicolumn{2}{c|}{$0.09$} &  \multicolumn{4}{c}{0.12} \\
 \hline
 \hline
op & op1  & op3 & op1  & op2 & op3 & op4 \\
\hline
\hline
$e^{i\xi_1}$ & $1A^{\prime\prime}$  & $3A^{\prime\prime}$  & $1A$  & 1.0 +53.3 i  & $3A$  & -0.000201 -1.0 i  \\
\hline
\hline
$e^{i\xi_2}$  &  $1A^-$    & $3A^-$     & $1A^\prime$   & 0.139 +0.990 i   & $3A^\prime$ &  0.936 +0.351 i \\
\hline
    \end{tabular}
    \end{center}
    \caption{$e^{i\xi_1}$ and $e^{i\xi_2}$}\label{ww}
\end{table}
Eq.~(\ref{cosxi2}) has been solved numerically and the results are given in Appendix \ref{appb}. It is straightforward to obtain $e^{i\xi_1}$ and $e^{i\xi_2}$ \footnote {since |$w_i$|=1, Cos[$\xi_i$]=rwi/gyi, gyi=$(rwi^2+iwi^2)^{1/2}$ is the normalization constant.} to the point. The results are listed in TABLE \ref{ww}. In TABLE \ref{ww}, the multiple solutions are as follows:
{\small
\bea
  1A &=& (1a,~1b,~1c,~1d,~1e)\nonumber\\
   &=& (-0.219 -0.976~i,~1.0 +11.467~i,~-0.998 +0.0650~i,~-0.998 -0.0645~i,~-0.259 +0.966~i)~,\nonumber\\
  1A^\prime &=& (1a^\prime,~1b^\prime,~1c^\prime,~1d^\prime,~1e^\prime)\nonumber\\
  &=& (-0.823 +0.568~i,~-0.707 -0.707~i,~-0.0538 -0.999~i,~0.0617 -0.998~i,~1.000 +0.00115~ i)~,\nonumber\\
  1A^{\prime\prime} &=& (1a^{\prime\prime},~1b^{\prime\prime},~1c^{\prime\prime},~1d^{\prime\prime},~1e^{\prime\prime})\nonumber\\
  &=& (-0.51 - 0.860~i,~-1.840\times 10^{-7} + 1.000~i,~-1.000 - 0.030~i,~-0.999 + 0.035~i,~-0.460 + 0.888~i)~,\nonumber\\
  1A^- &=& (1a^-,~1b^-,~1c^-,~1d^-,~1e^-)\nonumber\\
  &=& (-0.687 + 0.727~i,~-0.860 - 0.510~i,~-0.074 - 0.997~i,~0.085 - 0.996422~i,~-1.000+ 0.00301 i)~,\nonumber\\
  3A &=& (3a,~3b,~3c,~3d,~3e)\nonumber\\
  &=& (-0.305 -0.952~i,~1.0 +51.677~i,~-0.945 -0.327~i,~-0.937 +0.349~i,~-0.292 -0.956~ i)~,\nonumber\\
  3A^\prime &=& (3a^\prime,~3b^\prime,~3c^\prime,~3d^\prime,~3e^\prime)\nonumber\\
  &=& (-0.822 +0.570~i,~-0.170 +0.986~i,~-0.0539 -0.999~i,~0.0618 -0.998~i,~1.000 +0.00118 ~i)~,\nonumber\\
  3A^{\prime\prime} &=& (3a^{\prime\prime},~3b^{\prime\prime},~3c^{\prime\prime},~3d^{\prime\prime},~3e^{\prime\prime})\nonumber\\
  &=& (-0.51087 - 0.860 ~i,~0.0000146 - 1.000~i,~-1- 0.031~i,~-0.999 + 0.035~i,~-0.461 + 0.888~i)~,\nonumber\\
  3A^- &=& (3a^-,~3b^-,~3c^-,~3d^-,~3e^-) \nonumber\\
  &=& (-0.687 + 0.727~i,~-0.860 - 0.510~i,~-0.0744 - 0.997~i,~0.0846 - 0.996~i,~1.000 - 0.003 ~i)~.\nonumber\\
\eea
It is straightforward to compute the Majorana CP-violating phases and the effective Majorana neutrino masses $m_{ee}$. Results are listed in TABLE \ref{cpv} and TABLE \ref{cpv2}, respectively.
\nopagebreak

\begin{table}[htb]
\begin{center}
\begin{tabular}{ccc|ccccc}
\hline
$ sm /$eV &  \multicolumn{2}{c|}{$0.09$}    &   \multicolumn{4}{c}{$0.12$}   \\
\hline
\hline
op & op1 ($1A^{\prime\prime}$ or $1A^-$)  & op3 ($3A^{\prime\prime}$ or $3A^-$) & op1 ($1A$ or $1A^\prime$)  & op2 & op3 ($3A$ or $3A^\prime$) & op4 \\
\hline
\hline
$\xi_1 /^\circ$ &~120.65   &~120.72    &~ 102.62   &~ 0.00  &~ 107.77    &~ 90.01  \\
\hline
$\xi_1 /^\circ$ &~ 90.00    &~90.00    &~0.00    &~  &~0.00     &~ \\
\hline
$\xi_1 /^\circ$ &~178.26    &~178.25    &~176.28   &~  &~160.94   &~  \\
\hline
$\xi_1 /^\circ$ &~177.99    &~177.98    &~176.30  &~  &~159.57   &~  \\
\hline
$\xi_1 /^\circ$ &~117.40    &~117.42    &~105.04    &~  &~106.97   &~  \\
\hline
\hline
$\xi_2/^\circ$  &~133.39     &~133.39    &~145.42    &~81.99  &~145.28    &~20.58  \\
\hline
$\xi_2/^\circ$  &~149.32   &~149.32   &~134.99   &~     &~99.76      &~ \\
\hline
$\xi_2/^\circ$  &~94.26   &~94.27   &~93.08     &~   &~93.09      &~ \\
\hline
$\xi_2/^\circ$  &~85.15   &~85.15   &~86.46     &~   &~86.46      &~ \\
\hline
$\xi_2/^\circ$  &~0.18   &~0.18   &~ 0.08    &~   &~0.08      &~ \\
\hline
\end{tabular}
\end{center}
\caption{The Majorana CP-violating phases in the second model}\label{cpv}
\end{table}

\begin{table}[htb]
\begin{center}
{\footnotesize
\begin{tabular}{ccc|ccccc}
  \hline
 $ sm /$eV &  \multicolumn{2}{c|}{$0.09$}    &   \multicolumn{4}{c}{$0.12$}   \\
\hline
\hline
op & op1 ($1A^{\prime\prime}$ or $1A^-$)  & op3 ($3A^{\prime\prime}$ or $3A^-$) & op1 ($1A$ or $1A^\prime$)  & op2 & op3 ($3A$ or $3A^\prime$) & op4 \\
\hline
\hline
$m_{ee}/(0.01~\text{eV})$ & 0.898+0.657 i   & -1.024-0.485 i & 1.100+1.535 i &~3.741+1.574 i  & 1.414-1.296 i  & 1.438-2.990 i  \\
\hline
$|m_{ee}|/(0.01~\text{eV})$  &~1.113   &~1.133   &~1.888   &~4.059   &~1.918   &~3.317  \\
\hline
\hline
$m_{ee}/(0.01~\text{eV})$ &~-0.393+0.837 i &~0.517-1.366 i &~1.508+0.0719 i  &~ &~1.871+1.045 i &~  \\
\hline
$|m_{ee}|/(0.01~\text{eV})$  &~0.925   &~1.460    &~1.671    &~  &~2.143    &~  \\
\hline
\hline
$m_{ee}/(0.01~\text{eV})$ &~1.126+0.668 i &~-1.250-0.496 i &~-1.983-0.857 i &~ &~-2.012 -1.471 i  &~ \\
\hline
$|m_{ee}|/(0.01~\text{eV})$ &~1.309   &~1.345    &~2.160  &~ &~2.493     &~ \\
\hline
\hline
$m_{ee}/(0.01~\text{eV})$ &~1.034-0.589 i &~-1.158-0.417 i &~ 1.875+1.122 i  &~ &~-1.888-0.0848 i  &~ \\
\hline
$|m_{ee}|/(0.01~\text{eV})$ &~1.190   &~1.231    &~2.185   &~ &~1.890     &~ \\
\hline
\hline
$m_{ee}/(0.01~\text{eV})$ &~0.135+0.997 i &~0.0103+1.168 i &~0.513+1.921 i  &~ &~0.309-1.833 i &~ \\
\hline
$|m_{ee}|/(0.01~\text{eV})$ &~1.006   &~1.168    &~1.989  &~ &~1.859     &~ \\
\hline
    \end{tabular}}
    \end{center}
    \caption{The effective Majorana neutrino mass in the second model}\label{cpv2}
\end{table}

\vskip 2.0 cm
From the TABLE \ref{cpv2}, predicted $|m_{ee}|/(0.01~\text{eV})$  is smaller than or equal to 2.493 in all cases and is smaller than the upper limit, 0.061-0.165, obtained  in the experiments \cite{agando}.

\section{Summary and Outlook}\label{sec5}

In summary, we have proposed two phenomenological models, the first model in which we suppose the neutrino mass matrix is magical but the mixing matrix is experimentally measured, and the second model based on the assumption about the symmetry of the neutrino mass matrix that the neutrino mass matrix is both magic and $\mu-\tau$ reflect symmetric, in order to determine two Majorana CP-violating phases. We have computed the Majorana CP-violating
phases and the effective Majorana neutrino mass $m_{ee}$ in the two models. Using the nowadays data in experiments on neutrino oscillations only, the numerical values of the Majorana CP-violating phases are obtained in the first model. In the second model both the data on neutrino oscillations and the limit of the absolute neutrino mass scale which is from cosmological observations are needed in order to obtain the numerical values of the Majorana CP-violating phases. We have computed the numerical value of the effective Majorana neutrino mass $ m_{ee}$ which is in agree with the upper limit, 0.061 - 0.165 eV, obtained  in the neutrinoless double beta decay ($0\nu 2\beta$) experiments. The improved upper bound on the effective neutrino mass $m_{\beta}$ for $\beta$ decay experiments performed at KATRIN is reported to be smaller than $0.8$ eV at 90\% confidence level and future experiments for the effective Majorana neutrino mass $ m_{ee}$ hope to reach a goal of 10 - 40 meV\,\cite{kat,dpr}. Thus it is expected that the absolute neutrino mass scale shall be determined more accurately and the constraint to the Majorana CP-violating phases will be more tighter by experiments in the near future.

Although the absolute neutrino mass scale and the Majorana CP-violating phases are completely insensitive to normal neutrino oscillation experiments, they can be determined or constrained with the help of some non-oscillation processes, such as the beta
decays, neutrinoless double-beta decays ($0\nu 2\beta$)  and cosmological observations\,\cite{zxz}. However, as pointed in Ref.~\cite{bty} ten years ago, the Majorana CP phases are only sensitive to those lepton-number-violating processes which have never been measured since the lepton-number violating neutrino-antineutrino oscillations  cannot be observed in any realistic reactor experiments simply because their probabilities are typically suppressed by the tiny factors $m_i^2/E^2$ with $m_i$ being the neutrino mass and $E$ being the neutrino beam energy \cite{bty}. For instance, $m_i^2/E^2 \lesssim {\cal O} (10^{-14})$ is expected for the reactor antineutrino beam with $E \sim {\cal O} (1)$ MeV and $m_i \lesssim 0.1$ eV. Nevertheless, as pointed in Ref.~\cite{zzx}, one may consider to invent some new techniques and produce a sufficiently low energy neutrino (or antineutrino) beam. For example, the possibility of producing a $M\ddot{o}ssbauer$ electron antineutrino beam with $E$=18.6 keV \,\cite{vks} and using it to do an $\overline{\nu}_e \rightarrow \overline{\nu}_e$ oscillation experiment has been discussed \,\cite{mumn}. If the $\overline{\nu}_e \rightarrow \nu_e$ oscillation is taken into account in this case, the helicity suppression can be improved by a factor of $O(10^4)$ as compared with the case of the aforementioned reactor antineutrinos. In fact, as experimental techniques improve, in recent years  lepton flavor violation (LFV) processes such as $l_i \rightarrow l_j \gamma\gamma$ and $l_i \rightarrow l_j \gamma$ have been measured by experiments but  the upper limit on BR$(90\%~\text{c.l.})$ is $10^{-8}-10^{-13}$ for $l_i \rightarrow l_j \gamma$ and $10^{-4}-10^{-11}$ for $l_i \rightarrow l_j \gamma\gamma$ only\,\cite{lfv}. It is noted\,\cite{fim} that  in the frame of  an effective field theory  analysis, using the current upper bounds on the rate for $l_i \rightarrow l_j \gamma$, model-independent upper limits on the rates for $l_i \rightarrow l_j \gamma\gamma$ are derived and results indicate that indirect limits are about three orders of magnitude stronger than the direct bounds from current searches for the rare decay $\mu\rightarrow e \gamma\gamma$, and four orders of magnitude better than current bounds for $\tau \rightarrow l \gamma \gamma$. Furthermore, these limits do not preclude the possibility of observing the rare decays $\tau\rightarrow \mu \gamma \gamma$ at Belle II or at a Super Tau Charm Factory, and could constitute an stringent probe of lepton flavor violation.  Future seeable sensitivities of MEG II searching for $\mu\rightarrow e \gamma$  and of Belle II for $\tau\rightarrow l \gamma$ will improve the indirect limits by about one order of magnitude. As shown in the paper\,\cite{zzx} that there are the significant contributions of the Majorana phases to the CP-violating asymmetries of neutrino-antineutrino oscillations, even in the absence of the Dirac phase. Some new techniques for producing and measuring neutrinos and antineutrinos will probably be developed in the future\,\cite{df} and then the Majorana CP-violating phases will be measured. Therefore, it is look forward to measure the Majorana CP-violating phases experimentally in the future so to verify the models in the paper.

\section*{Acknowledgments}

This research was supported in part by the National Natural Science Foundation of China under grants No. 11875306, No. 11875062, and No. 12275335.

\appendix
\section{Expressions of $A_i$, $B_i~(i=0,1,2)$ and $C_j$, $D_j~(j=0,1,...,4)$}\label{appa}
\bea
A_0 &=& 16 c_{12}^2 (m_1 + m_2 + m_3)^2 s_{12}^2 s_{13}^2, \nnb\\
A_1 &=& -8 \sqrt{6}c_{12}^2 c_{13} (m_2 - m_3)(m_1 +m_2+m_3)s_{12} s_{13},
\nnb\\
A_2 &=& 6 c_{12}^2 c_{13}^2 (m_2 - m_3)^2, \nnb\\
B_0 &=& 16 c_{12}^2 (m_1 + m_2 + m_3)^2 s_{12}^2 s_{13}^2, \nnb \\
B_1 &=& -8 \sqrt{6}c_{12}^2 c_{13} (m_2 - m_3) (m_1 + m_2 + m_3) s_{12} s_{13},\nnb\\
B_2 &=& 6 c_{12}^2 c_{13}^2 (m_2 - m_3)^2, \nnb\\
C_0 &=& 48 c_{13}^2 (m_2-m_3)^2 (m_1+m_2+m_3)^2 (2 m_1^2+m_1 (m_2+m_3)-(m_2+m_3)^2)^2 s_{12}^2  (c_{12}^2+s_{12}^2 s_{13}^2)^2 \nnb\\ && (c_{12}^2 (-c_{13}^2+s_{13}^2)+s_{12}^2 s_{13}^2 )^2, \nnb \\
C_1 &=& -48 \sqrt{6}c_{13}^3 (m_2 - m_3)^3 (m_1 + m_2 + m_3)^2 (2 m_1^3 - 3 m_1^2 (m_2 + m_3) -3 m_1 (m_2 + m_3)^2 +  2 (m_2 + m_3)^3) \nnb\\
&&s_{12}^7 s_{13}^3 (c_{12}^2 +s_{12}^2 s_{13}^2) (s_{12}^2 s_{13}^2 + c_{12}^2 (-c_{13}^4 + s_{13}^4)), \nnb\\
C_2 &=& (m_2 - m_3)^2 ((-27 + 4 \sqrt{3}) m_1^2 + 2 (-27 + \sqrt{3}) (m_1 - m_2 - m_3) (m_2 + m_3)) s_{12}^{10}
s_{13}^6 - 2 (-9 + \sqrt{3}) c_{12}^2 \nnb \\
&&(2 m_1 - m_2 - m_3) (m_1 + m_2 + m_3) (c_{12}^8 (c_{13}^2 - s_{13}^2) (c_{13}^2 (2 m_1 - m_2 - m_3) (m_2 + m_3) + (m_2 - m_3)^2 s_{13}^2) \nnb \\
&&+ c_{12}^6 s_{12}^2 ((m_2 - m_3)^2 s_{13}^4 (2 + 3 s_{13}^2) - c_{13}^4  (4 m_1 (m_2 + m_3) + m_2^2 (-2 + s_{13}^2) + m_3^2 (-2 + s_{13}^2) - 2 m_2 m_3 \nnb \\
&&(2 + s_{13}^2)) + 2 c_{13}^2 s_{13}^2 (-2 m_2 m_3 s_{13}^2 + 2 m_1 (m_2 + m_3) (1 + s_{13}^2) - m_2^2 (2 + s_{13}^2) - m_3^2 (2 + s_{13}^2))) + s_{12}^8 s_{13}^4\nnb \\
&&((m_2-m_3)^2 s_{13}^2 (3 + 2 s_{13}^2) + c_{13}^2 (2 m_1 (m_2 + m_3) + m_2^2 (-1 + 2 s_{13}^2) + m_3^2 (-1 + 2 s_{13}^2) - 2 m_2 (m_3 + 2 m_3 s_{13}^2)))\nnb \\
&& + c_{12}^4 s_{12}^4 ((m_2 - m_3)^2 s_{13}^4 (1 + 6 s_{13}^2 + 3 s_{13}^4) + c_{13}^4 (2 m_1 (m_2 + m_3) (-1 +  s_{13}^4) - 2 m_2 m_3 (-1 - 2 s_{13}^2 + s_{13}^4)\nnb \\
&& - m_2^2 (-1 + 2 s_{13}^2 + s_{13}^4) - m_3^2 (-1 + 2 s_{13}^2 +  s_{13}^4)) + 2 c_{13}^2 s_{13}^2 (-4 m_2 m_3 s_{13}^2 (1 + s_{13}^2) + m_2^2 (-1 - 2 s_{13}^2 + s_{13}^4) \nnb \\
&&+ m_3^2 (-1 - 2 s_{13}^2 + s_{13}^4) + m_1 (m_2 + m_3) (1 + 4 s_{13}^2 + s_{13}^4))) + c_{12}^2 s_{12}^6 s_{13}^2 ((m_2 - m_3)^2 s_{13}^4 (3 + 6 s_{13}^2 + s_{13}^4) \nnb \\
&&+ 2 c_{13}^2 s_{13}^2 (2 m_1 (m_2 + m_3) (1 + s_{13}^2) + m_2^2 (-1 + 2 s_{13}^2 + s_{13}^4) + m_3^2(-1 + 2 s_{13}^2 + s_{13}^4) - 2 m_2 m_3(1 + 4 s_{13}^2 \nnb \\
&&+ s_{13}^4)) + c_{13}^4 (4 m_1 m_2 s_{13}^2 + m_2^2 (-1 - 2 s_{13}^2 + s_{13}^4) - 2 m_2 m_3 (-1 + 2 s_{13}^2 + s_{13}^4) + m_3 (4 m_1 s_{13}^2 + m_3 (-1 \nnb \\
&&- 2 s_{13}^2 + s_{13}^4))))) , \nnb \\
C_3 &=& -24 \sqrt{2} (-1 + 3 \sqrt{3}) c_{13}^5 (m_2 - m_3)^3 (m_1 + m_2 + m_3)^2 (-m_1 + 2 (m_2 + m_3)) s_{12}^7  s_{13}^3 ((m_2 - m_3)^2 s_{12}^2 s_{13}^2 \nnb \\
&&+ c_{12}^2 ((m_2 - m_3)^2 - 2 c_{13}^2 (m_2^2 + m_3^2 - m_1 (m_2 + m_3)))) , \nnb\\
C_4 &=& -8 (-14 + 3 \sqrt{3}) c_{13}^6 (m_2 - m_3)^2 (m_1 + m_2 + m_3)^2 s_{12}^2 ((m_2 - m_3)^2 s_{13}^2 \nnb \\
&&+ c_{12}^2 (m_2 + m_3) (-2 m_1 + m_2 + m_3) (- c_{13}^2))^2 , \nnb\\
D_0 &=& 48 c_{13}^2 (2 m_1 - m_2 - m_3) (m_2 - m_3)^2 (m_1 + m_2 + m_3)^3 (2 m_1^2 + (m_1 - m_2 - m_3) (m_2 + m_3)) s_{12}^2 (c_{12}^2 + s_{12}^2 s_{13}^2)^4 , \nnb\\
D_1 &=& 48 \sqrt{6} c_{13}^3 (2 m_1 - m_2 - m_3) (m_2 - m_3) (m_1 + m_2 + m_3)^3 s_{12}^3 s_{13} (c_{12}^2 + s_{12}^2 s_{13}^2)^2 (2 c_{12}^4 c_{13}^2 \nnb\\
&&(m_2 + m_3)(-2 m_1^2 + (m_2 + m_3) (-m_1 + m_2 + m_3)) - (m_2 - m_3)^2 (m_1 - 2 (m_2 + m_3)) s_{12}^4 s_{13}^2 \nnb\\
&&+ 2 c_{12}^2 (m_2 - m_3)^2 s_{12}^2 s_{13}^2 (m_1 (-2 + s_{13}^2) + (m_2 + m_3) (1 + s_{13}^2)) + c_{12}^4 (m_2 - m_3)^2 (2 (m_2 + m_3) s_{13}^2 \nnb\\
&&+ m_1 (-3 + 2 s_{13}^2)) + c_{12}^2 c_{13}^2 s_{12}^2 (-4 m_1^2 (m_2 + m_3) + 2 (m_2 + m_3) (-2 m_2 m_3 (-1 + s_{13}^2) +
(m_2^2 + m_3^2)\nnb\\
&& (1 + s_{13}^2)) + m_1 (m_2 m_3 (2 - 4 s_{13}^2) + (m_2^2 + m_3^2) (-5 + 2 s_{13}^2)))) , \nnb\\
D_2 &=& 72 c_{13}^4 (m_1 + m_2 + m_3)^2 s_{12}^2 (-3 (m_2 - m_3)^4 (m_1^2 + 2 m_1 (m_2 + m_3) - 2 (m_2 + m_3)^2) s_{12}^{10} s_{13}^6 + \nnb\\
&&2 c_{12}^{10} (m_2 - m_3)^2 (2 m_1^2 + m_1 (m_2 + m_3) - (m_2 + m_3)^2) (c_{13}^2 (2 m_1 - m_2 - m_3) (m_2 + m_3)  - \nnb\\
&&(m_2 - m_3)^2 s_{13}^2) + 2 c_{12}^2 (m_2 - m_3)^2 s_{12}^8 s_{13}^4 (m_1 (m_2 - m_3)^2 s_{13}^2 (-13 (m_2 + m_3) - 2 m_1 (1 + 3 s_{13}^2)) + \nnb\\
&&c_{13}^2  (8 m_1^3 (m_2 + m_3) -
 m_1 (17 m_2^3 + 19m_2^2 m_3 + 19 m_2 m_3^2 + 17 m_3^3) + 6 m_2 m_3 (m_2 + m_3)^2 (1 + \nnb\\
 &&(m_2 + m_3)^2) (1 - 2 s_{13}^2) + 2 m_1^2 m_2 m_3
(-5 + 6 s_{13}^2))) c_{12}^8 s_{12}^2 (4 c_{13}^2 (m_2 - m_3)^2 (-2 m_1^2 - m_1 (m_2 + m_3) \nnb\\
&&+ (m_2 + m_3)^2) s_{13}^2 (-5 m_1 (m_2 + m_3) +
 2 m_2 m_3 s_{13}^2 (m_2^2 + m_3^2) (2 + 5 s_{13}^2)) + 2 c_{13}^4 (2 m_1^2 + m_1 (m_2 \nnb\\
&&+ m_3) - (m_2 + m_3)^2) (-4 m_2^3 m_3 s_{13}^2 - 4 m_2 m_3^3 s_{13}^2 + 4 m_1^2(m_2 + m_3)^2 s_{13}^2 + 2 m_2^2 m_3^2 (2 - 7 s_{13}^2) \nnb\\
&&+ (m_2^2 + m_3^2) (2 + 3 s_{13}^2) - (m_2^4 + m_3^4) (2 + 5 s_{13}^2)) + (m_2 - m_3)^4 s_{13}^2 (2 (m_2 + m_3)^2 s_{13}^2 (2 + 5 s_{13}^2) \nnb\\
&&+ m_1 (2 (m_2 + m_3) s_{13}^2 (-8 + s_{13}^2) + m_1 (9 - 20 s_{13}^2 - 8 s_{13}^4)))) + c_{12}^4 s_{12}^6 s_{13}^2 (2 (m_2 - m_3)^4 s_{13}^4 \nnb\\
&&(m_1^2 (5 - 22 s_{13}^2) + m_1 (m_2 + m_3) (-17 - 8 s_{13}^2 + 3 s_{13}^4)
+ (m_2 + m_3)^2 (5 + 14 s_{13}^2 + 3 s_{13}^4)) + \nnb\\
&&2 c_{13}^2 (m_2 - m_3)^2 s_{13}^2 (4 m_1^3 (m_2 + m_3) (6 + s_{13}^2) + m_1^2 ((m_2^2 + m_3^2) (7 - 54 s_{13}^2) +2 m_2 m_3 (-7 + \nnb\\
&&18 s_{13}^2)) + 2 m_1 ((m_2^3 + m_3^3) (-17 - 15 s_{13}^2 + 3 s_{13}^4) - m_2 m_3 (m_2 + m_3) (19 + 9 s_{13}^2 + 3 s_{13}^4)) +\nnb\\
&& 2 (m_2 + m_3)^2 (-2 m_2 m_3 (-1 + 4 s_{13}^2 + 3 s_{13}^4) + (m_2^2 + m_3^2) (5 + 14 s_{13}^2 + 3 s_{13}^4))) + c_{13}^4 (16 m_1^4 \nnb\\
&&(m_2 + m_3)^2 + 8 m_1^3 (m_2 + m_3) (-2 m_2 m_3 (3 + s_{13}^2) + (m_2^2 + m_3^2) (7 + s_{13}^2)) - m_1^2 (6 m_2^2 m_3^2 \nnb\\
&&(11 + 16 s_{13}^2) - 4 m_2 m_3 (m_2^2 + m_3^2) (-17 + 28 s_{13}^2) + (m_2^4 + m_3^4) (-5 + 64 s_{13}^2)) + 2 m_1 (m_2 + m_3) \nnb\\
&&((m_2^4 + m_3^4) (-17 - 22 s_{13}^2 + 3 s_{13}^4) - 4 m_2 m_3 (m_2^2 + m_3^2) (3 - 10 s_{13}^2 + 3 s_{13}^4) + 6 m_2^2 m_3^2 (-1 \nnb\\
&&- 6 s_{13}^2 + 3 s_{13}^4)) + 2 (m_2 + m_3)^2 (-4 m_2 m_3 (m_2^2 + m_3^2) (-1 + 4 s_{13}^2 + 3 s_{13}^4) + (m_2^4 + m_3^4) (5 + \nnb\\
&&14 s_{13}^2 + 3 s_{13}^4) + 2 m_2^2 m_3^2 (7 + 2 s_{13}^2 + 9 s_{13}^4)))) + 2 c_{12}^12 s_{12}^4 ((m_2 - m_3)^4 s_{13}^4 (-2 m_1^2 (-5 + 13 s_{13}^2 \nnb\\
&&+ s_{13}^4) + m_1 (m_2 + m_3) (-7 - 16 s_{13}^2 + 5 s_{13}^4) + (m_2 + m_3)^2 (1 + 10 s_{13}^2 + 7 s_{13}^4)) + c_{13}^4 (16 m_1^4 \nnb\\
&&(m_2 + m_3)^2 s_{13}^2 - 2 m_1^2 s_{13}^2 (54 m_2^2 m_3^2 - 8 m_2 m_3 (m_2^2 + m_3^2) (-2 + s_{13}^2) + (m_2^4 + m_3^4) (5 + 8 s_{13}^2)) \nnb\\
&&- 4 m_1^3 ((m_2^3 + m_3^3) (-1 - 11 s_{13}^2 + s_{13}^4) - m_2 m_3 (m_2 + m_3) (-1 + 5 s_{13}^2 + s_{13}^4)) - m_1 (m_2 + m_3) \nnb\\
&&(-8 m_2 m_3 (m_2^2 + m_3^2) s_{13}^2 (-3 + s_{13}^2) + 6 m_2^2 m_3^2 (-1 + 4 s_{13}^2 + s_{13}^4) + (m_2^4 + m_3^4) (3 + 28 s_{13}^2 + \nnb\\
&&5 s_{13}^4)) + (m_2 + m_3)^2 (-8 m_2 m_3 (m_2^2 + m_3^2) s_{13}^2 (-1 + s_{13}^2) + 2 m_2^2 m_3^2 (-1 + 14 s_{13}^2 + s_{13}^4) +\nnb\\
&& (m_2^4 + m_3^4) (1 + 10 s_{13}^2 + 7 s_{13}^4))) + c_{13}^2 (m_2 - m_3)^2 s_{13}^2 (-4 m_1^3 (m_2 + m_3) (-4 - 6 s_{13}^2 + s_{13}^4) -2 m_1 \nnb\\
&&((m_2^3 + m_3^3) (5 + 22 s_{13}^2) + m_2 m_3 (m_2 + m_3) (13 + 26 s_{13}^2 + 6 s_{13}^4)) + 2 (m_2 + m_3)^2  (-4 m_2 m_3 s_{13}^2 \nnb\\
&&(-1 + s_{13}^2) + (m_2^2 + m_3^2) (1 + 10 s_{13}^2 + 7 s_{13}^4)) - m_1^2 (2 m_2 m_3 (1 - 16 s_{13}^2 + 6 s_{13}^4) + (m_2^2 + m_3^2)\nnb\\
&&(-13 + 40 s_{13}^2 + 18 s_{13}^4))))) , \nnb\\
D_3 &=& -72 \sqrt{6} c_{13}^5 (m_2 - m_3) (m_1 + m_2 + m_3)^2 s_{12}^3 s_{13} (-(m_2 - m_3)^4 (m_1 - 2 (m_2 + m_3)) s_{12}^6 s_{13}^4 + c_{12}^6 (2 c_{13}^4 \nnb\\
&&(m_2 + m_3)^2 (-2 m_1 + m_2 + m_3)^2 (m_1 + m_2 + m_3) + (m_2 - m_3)^4 s_{13}^2 (2 (m_2 + m_3) s_{13}^2 + m_1 (-3 + 2 s_{13}^2))\nnb\\
&&+ c_{13}^2 (m_2 - m_3)^2 (m_2 + m_3) (4 (m_2 + m_3)^2 s_{13}^2 + m_1^2 (6 - 8 s_{13}^2) - m_1 (m_2 + m_3) (3 + 4 s_{13}^2))) + c_{12}^2 \nnb\\
&&(m_2 - m_3)^2 s_{12}^4 s_{13}^2 ((m_2 - m_3)^2 s_{13}^2 (2 (m_2 + m_3) (2 + s_{13}^2) + m_1 (-5 + 2 s_{13}^2)) - 2 c_{13}^2 (m_1^2 (m_2 + m_3) \nnb\\
&&+ (2 m_1 m_2 m_3- (m_2 + m_3) (m_2^2 + m_3^2)) (2 + s_{13}^2) + m_1 (-(m_2^2 + m_3^2) (-5 + s_{13}^2) + 2 m_2 m_3 (m_2 + m_3) \nnb\\
&&(-2 + s_{13}^2)))) + c_{12}^4 s_{12}^2 ((m_2 - m_3)^4 s_{13}^4 (2 (m_2 + m_3) (1 + 2 s_{13}^2) + m_1 (-7 + 4 s_{13}^2)) + c_{13}^4 (2 m_1 - m_2 \nnb\\
&&- m_3) (m_2 + m_3) (4 m_1^2 (m_2 + m_3) - 2 (m_2 + m_3) (m_2 m_3 (2 - 4 s_{13}^2) + (m_2^2 + m_3^2) (1 + 2 s_{13}^2)) + m_1 ((\nnb\\
&&m_2^2 + m_3^2) (5 - 4 s_{13}^2) + 2 m_2 m_3 (-1 + 4 s_{13}^2))) + 4 c_{13}^2 (m_2 - m_3)^2 s_{13}^2 (-m_1^2 (m_2 + m_3) (-1 + 2 s_{13}^2) \nnb\\
&&+ (m_2 + m_3) (2 m_2 m_3 + (m_2^2 + m_3^2) (1 + 2 s_{13}^2)) - 2 m_1 (2 m_2^2 + 2 m_3^2 + m_2 (m_3 + 2 m_3 s_{13}^2))))) , \nnb\\
D_4 &=& -72 \sqrt{6} c_{13}^5 (m_2 - m_3) (m_1 + m_2 + m_3)^2 s_{12}^3 s_{13} (-(m_2 - m_3)^4 (m_1 - 2 (m_2 + m_3)) s_{12}^6 s_{13}^4 + c_{12}^6 (2 c_{13}^4 \nnb\\
&&(m_2 + m_3)^2 (-2 m_1 + m_2 + m_3)^2 (m_1 + m_2 + m_3) + (m_2 - m_3)^4 s_{13}^2 (2 (m_2 + m_3) s_{13}^2 + m_1 (-3 + 2 s_{13}^2))\nnb\\
&&+ c_{13}^2 (m_2 - m_3)^2 (m_2 + m_3) (4(m_2 + m_3)^2 s_{13}^2 + m_1^2 (6 - 8 s_{13}^2) - m_1 (m_2 + m_3) (3 + 4 s_{13}^2))) + c_{12}^2 \nnb\\
&&(m_2 - m_3)^2 s_{12}^4 s_{13}^2 ((m_2 - m_3)^2 s_{13}^2 (2 (m_2 + m_3) (2 + s_{13}^2) + m_1 (-5 + 2 s_{13}^2)) - 2 c_{13}^2 (m_1^2 (m_2 + m_3) \nnb\\
&&+ (2 m_1 m_2 m_3 - (m_2 + m_3)(m_2^2 + m_3^2)) (2 + s_{13}^2) + m_1 (- (m_2^2 + m_3^2) (-5 + s_{13}^2) + 2 m_2 m_3 (m_2 + m_3) \nnb\\
&&(-2 + s_{13}^2)))) + c_{12}^4 s_{12}^2 ((m_2 - m_3)^4 s_{13}^4 (2 (m_2 + m_3) (1 + 2 s_{13}^2) + m_1 (-7 + 4 s_{13}^2)) + c_{13}^4 (2 m_1 - \nnb\\
&&m_2 - m_3) (m_2 + m_3) (4 m_1^2 (m_2 + m_3) - 2 (m_2 + m_3) (m_2 m_3 (2 - 4 s_{13}^2) + (m_2^2 + m_3^2) (1 + 2 s_{13}^2)) + \nnb\\
&&m_1 ((m_2^2 + m_3^2) (5 - 4 s_{13}^2) + 2 m_2 m_3 (-1 + 4 s_{13}^2))) + 4 c_{13}^2 (m_2 - m_3)^2 s_{13}^2 (-m_1^2 (m_2 + m_3) (-1 + \nnb\\
&&2 s_{13}^2) + (m_2 + m_3) (2 m_2 m_3 + (m_2^2 + m_3^2) (1 + 2 s_{13}^2)) - 2 m_1 (2 m_2^2 + 2 m_3^2 + m_2 (m_3 + 2 m_3 s_{13}^2))))).  \nnb\\
\eea \nnb \\

\section{Solutions of rw2}\label{appb}

$sm$=0.12 eV:
\begin{itemize}
    \item op1 has five solutions rw2=-5.045,~-0.599,~-0.00626,~0.00683,~5.033.
    \item op2 has one solution rw2=0.431.
    \item op3 has five solutions rw2=-5.023,~-0.599,~-0.00627,~0.00684,~ 5.006.
    \item op4 has one solution rw2=0.431.
\end{itemize}

$sm$=0.09 eV:
\begin{itemize}
    \item op1 has five solutions rw2=3.384,~-0.363,~-0.00891,~0.00965,~3.375.
    \item op3 has five solutions rw2=-3.383,~-0.363,~-0.00892,~0.00966,~3.374.
\end{itemize}

\end{document}